# Development and space-qualification of a miniaturized CubeSat's 2-W EDFA for space laser communications

Alberto Carrasco-Casado [1,*], Koichi Shiratama [1], Dimitar Kolev [1], Phuc V. Trinh [1], Femi Ishola [1], Tetsuharu Fuse [1], Morio Toyoshima [2]

[1] National Institute of Information and Communications Technology (NICT),
Space Communication Systems Laboratory, 184-8795 Tokyo, Japan

[2] National Institute of Information and Communications Technology (NICT),
Wireless Networks Research Center, 184-8795 Tokyo, Japan

* Correspondence: alberto@nict.go.jp

**Abstract:** The Japanese National Institute of Information and Communications Technology (NICT) is currently developing a high-performance laser-communication terminal for CubeSats aiming at providing a high-datarate communication solution for LEO satellites requiring to transmit large volumes of data from the orbit. A key part of the communication system is a high-power optical amplifier capable of providing enough gain to the transmitted signals to be able to close the link on its counterpart's receiver with the smallest impact in terms of energy and power on the CubeSat's platform. This manuscript describes the development of a miniaturized 2-W space-grade 2-stage Erbium-Doped Fiber Amplifier (EDFA) compatible with the CubeSat form factor, showing the best power-to-size ratio for a spaced-qualified EDFA to the best of the authors' knowledge. Performance results under realistic conditions as well as full space qualification and test are presented, proving that this module can support short-duration LEO-ground downlinks as well as long-duration inter-satellite links.

**Keywords:** EDFA; CubeSat; satellite communication; free-space laser communications; optical amplifier; space qualification; thermal vacuum

## 1. Introduction

The National Institute of Information and Communications Technology (NICT) of Japan is one of the pioneers in the field of space optical communications. Since the 90's, NICT has carried out some of the most significant in-orbit demonstrations, paving the way for the generalized establishment of this technology in the communication networks [1]. Developing universal free-space laser communication (lasercom) terminals to fit the needs of all kinds of platforms is one of the current strategic plans of NICT. Within this endeavor, a series of versatile miniaturized terminals are presently being designed and developed, with the goal of demonstrating them in several different scenarios onboard several different platforms [2].

CubeSats in low-Earth orbit (LEO) is not only one of these, but conceivably the most important one from the miniaturization standpoint, because it pushes the size, weight and power (SWAP) requirements to the limit [3]. Following the path of the NICT's SOTA mission [4], the miniaturized lasercom terminal CubeSOTA was conceived to integrate in an ultra-small terminal all the capabilities of high-end space optical communication systems [5]. CubeSOTA aims at providing a high datarate communication solution for LEO satellites requiring to transmit large volumes of data from the orbit. Earth-observation satellites is a typical example of these users, who do not require to receive high-speed communications in the satellite, but only to download all the imaging data acquired by the spacecraft sensors.

A key part of this communication system is a high-power optical amplifier [6] [7] which can provide enough gain to the transmitted signals to be able to close the link on the ground's receiver with the smallest impact on the CubeSat's platform in terms of size, weight, and power, and the capability to survive the space environment [8]. This manuscript describes the development of a miniaturized 2-W space-grade 2-stage Erbium-Doped Fiber Amplifier (EDFA) compatible with the CubeSat form factor (fig. 1). Performance results under realistic conditions, based on the LEO-orbit environment, as well as its full space qualification and extensive test campaign are presented.

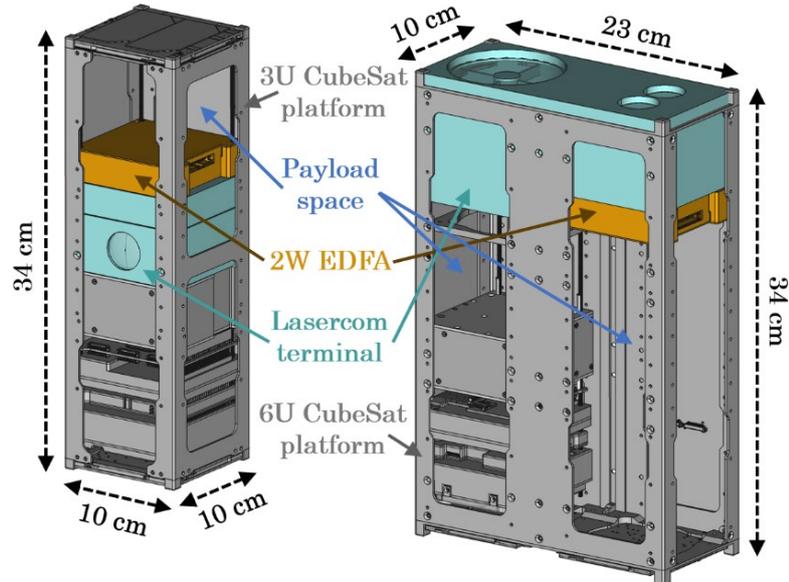

**Figure 1.** 3D model of CubeSOTA showing the EDFA within the lasercom terminal and its impact on the CubeSat's platforms.

## 2. EDFA's basic design and specifications

Since the EDFA must be compatible with the CubeSat's form factor, its footprint was designed to be 90 × 95 mm, and its height was optimized to fit in a case as small as 25 mm (fig. 2) with a total mass of 430 grams. The EDFA's interface is made up by power and control. The former requires one line of 3.3V and one of 5V, which are common in the CubeSat platform. For the latter, the EDFA's internal microcontroller is equipped with an UART (Universal Asynchronous Receiver-Transmitter) port to communicate with the CubeSat's onboard computer (OBC), although an UART-to-RS232 converter was developed for tests with PC.

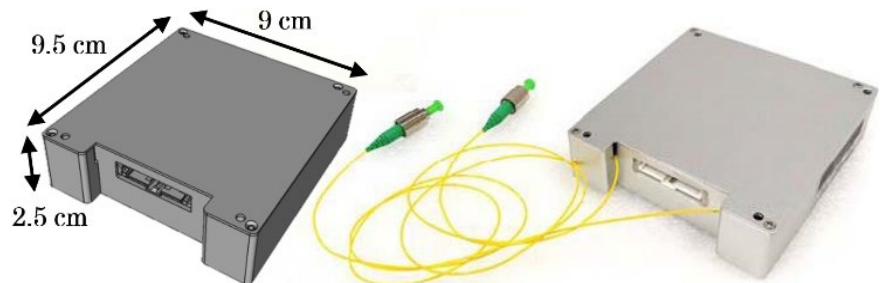

**Figure 2.** 2-W EDFA's 3D model (left) and flight model (right).

The physical CubeSat's interface is implemented with a 12-pin Molex connector to provide double-pin configuration for redundancy of the 5 interface lines: 3.3V, 5V, GND, UART tx and UART rx. The control of the EDFA allows to monitor the optical input power, optical output power and case temperature (via an onboard temperature sensor), as well as to set the EDFA's enable/disable and optical output power via driving current of the second pump laser (the first pump laser is set to the maximum current after EDFA has been enabled). As for the optical interface, the input and output single-mode fibers both have tight-buffered 900-μm fiber cables with FC/APC connectors in both ends. A small cavity was designed in one of the faces of the case in order to allow the fibers and the 12-pin cable go upwards or downwards within the CubeSat structure. The block diagram of the CubeSat's EDFA is shown in fig. 3. The diagram shows the dual-stage configuration which allows to achieve a better gain and noise figure combination by implementing low-noise preamplification in the first stage and high-power amplification in the second stage [9].

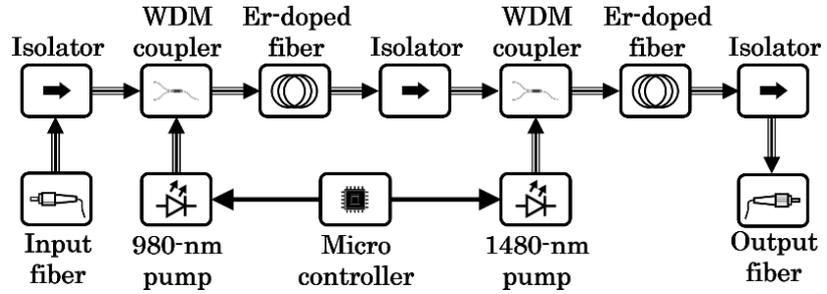

**Figure 3.** Block diagram of the 2-W CubeSat's EDFA.

The polarization-dependent gain is less than 0.5 dB and typically 0.3 dB, with an isolation between input and output bigger than 30 dB. This EDFA was designed to be independent to the laser's polarization state for various reasons. Firstly, polarization-maintaining (PM) fibers are more expensive, especially when required to be radiation-hardened, and this amplifier was conceived to be a low-cost device that facilitates the future adoption of the host lasercom terminal as a commodity when space laser communications are widely used. Additionally, a PM design would introduce supplementary losses, estimated to reach up to an extra 5% in the worst case. Furthermore, polarization-division multiplexing (PDM) is a good candidate for future lasercom systems based on coherent detection, and this technique requires non-PM components.

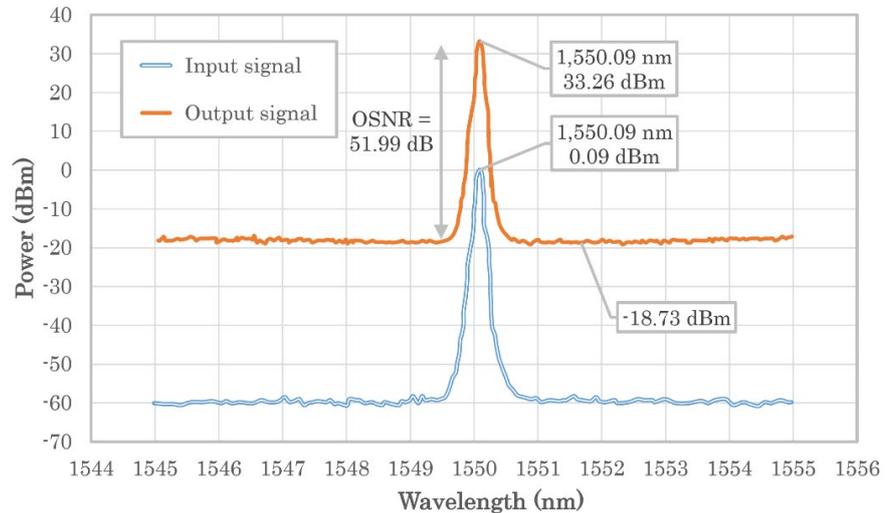

**Figure 4.** Spectrum-analyzer measurement of the 2-W EDFA.

Figure 4 shows the performance of the 2-W EDFA using an optical spectrum analyzer (OSA) for an input-signal's wavelength of 1550 nm. The OSA spectrum shows the 2 W level at 33.26 dBm and the noise due to the amplified spontaneous emission (ASE) at -18.73 dBm. The EDFA provides an equally approximated saturated gain when the input power is within -6 dBm and 3 dBm. However, the ASE noise is proportional to the input power, going from -15.2 dBm/0.1 nm at -6 dBm to -21.6 dBm/0.1 nm at 3 dBm. Therefore, the input power should be kept as small as possible within its tolerable range to optimize the optical signal-to-noise ratio (OSNR), which is 51.99 dB when input power is 0 dBm.

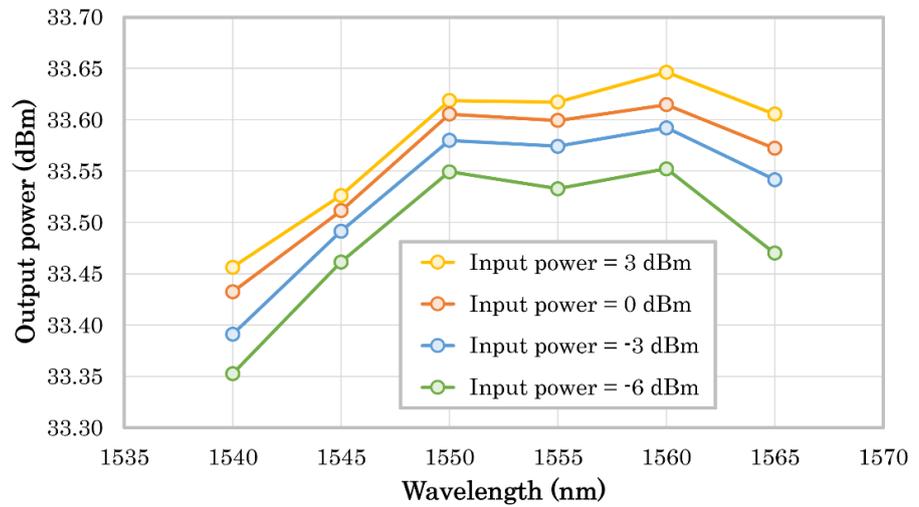

**Figure 5.** Dependency of the EDFA's output optical power with wavelength for different levels of input optical power.

Figure 5 shows the dependency of the output optical power with wavelength for different levels of input power. The maximum output power range is around 1550-1560 nm, and the variation with input power is typically around 0.1 dB within the whole range and smaller than 0.2 dB in the worst case. The output power can be controlled by one of the SCPI commands via the UART serial communications by adjusting the 2nd-stage pump laser driving current. The EDFA's control mode is based on an automatic pump-current control (ACC) to maintain a fixed pump laser current. Automatic power control (APC) may seem more appropriate since the important parameter in laser communication is the transmitted power. However, in a CubeSat there is a very severe restriction in the available power, thus the current is limited, and rather than maintaining a constant power, it is preferable to transmit the maximum available power, which can also be controlled indirectly through the current. Additionally, APC would require some added complexity for a tap splitter and a photodiode to monitor the power. When the EDFA is operating in the 2 W (full power) regime with saturated gain, the measured consumption of the device is around 18 W (see table 1 for more detailed consumption figures), taking 85% of the current from the 5 V line and 15% from the 3.3 V line.

### 3. EDFA's space qualification and test

The 2-W EDFA was conceived to be operated in LEO orbit within a CubeSat, which means that the exposure to the space environment can be even more intense compared with other conventional satellites. Therefore, the space qualification and test are critical steps before proceeding with the integration with the spacecraft prior to launch. The ISO 15864:2004 (Space systems — General test methods for space craft, subsystems and units) and ISO 19683:2017 (Space systems — Design qualification and acceptance tests of small spacecraft and units) standards were followed as a reference guide to comply with the qualification for sinusoidal and random vibration tests, thermal vacuum tests (TVT), and total ionization dose (TID) test. Three identical EDFA modules were produced, two of which went through the full qualification tests to certify that the third one is prepared for operation in the space environment. One of the units was used to make a preliminary investigation under partial high-temperature vacuum test prior to the complete TVT, as well as a preliminary TID in facilities with simpler equipment than the final one, which was the Center for Nanosatellite Testing (CeNT) of the Kyushu Institute of Technology in Japan.

*3.1. Radiation tests*

The basic strategy to protect the EDFA against radiation was to shield the electronics module with an Aluminum case instead of utilizing radiation-hardened components, with the exception of the fibers, which are partially exposed outside the case. To verify this approach, a preliminary test was carried out by using two dummy (nonfunctional) units which include only the electronic board inside the case (see fig. 6). In order to evaluate any variation after the TID test, the internal voltage and current were measured in 11 critical points of each of the 2 units, including the DC/DC converters, the low-dropout voltage regulators, and the pump-current drivers. Non-significant variations, in the order of 0.1%, were found before and after these tests.

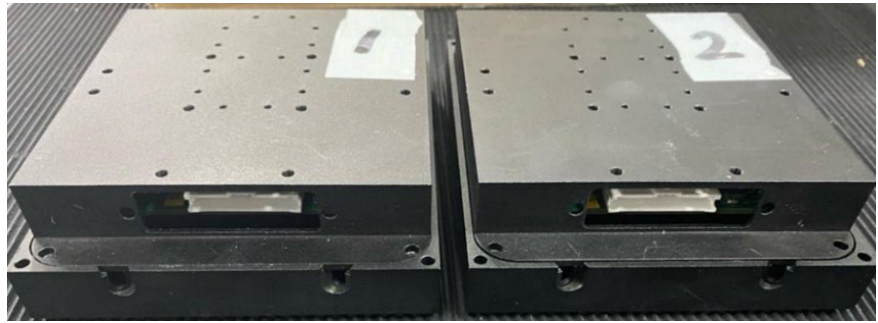

**Figure 6.** Dummy EDFA units for the preliminary radiation tests.

The final radiation tests were carried out over the final two test EDFA modules, which are identical to the third flight model, with the goal of verifying their overall performance. During these tests, the output optical power as well as the noise figure were measured before and after the radiation exposure for the stage 1, stage 2, and for the whole EDFA module. The total degradation was smaller than 0.1 dB for the optical power (from 32.44 to 32.36 dB) and smaller than 0.5 dB for the noise figure (from 5.47 to 5.94 dB) after being exposed to a gamma-radiation TID of 20 krad.

*3.2. Vibration tests*

The sinusoidal vibration tests consisted of a functional verification before and after applying a sine sweep within the range of 5-140 Hz (vibration amplitude of 2.5 g in the range of 5 100 Hz and 1.25 g in 100-140 Hz, with 1 g being the unit of measure of acceleration, equals to 9.80665 m/s$^2$ in the International System of Units) with 3 repetitions for each orthogonal axis. The functional tests before and after random vibration were performed during 2 minutes for each test with 3 repetitions for each orthogonal axes within the range of 20-2000 Hz. As shown in fig. 7, the vibration tests were done in two different vibration instruments, one for horizontal vibration (X-Y) and one for vertical vibration (Z). The output optical power was measured before and after each test with no significant variation that could be attributed to the vibration tests, in the order of 0.1 dB.

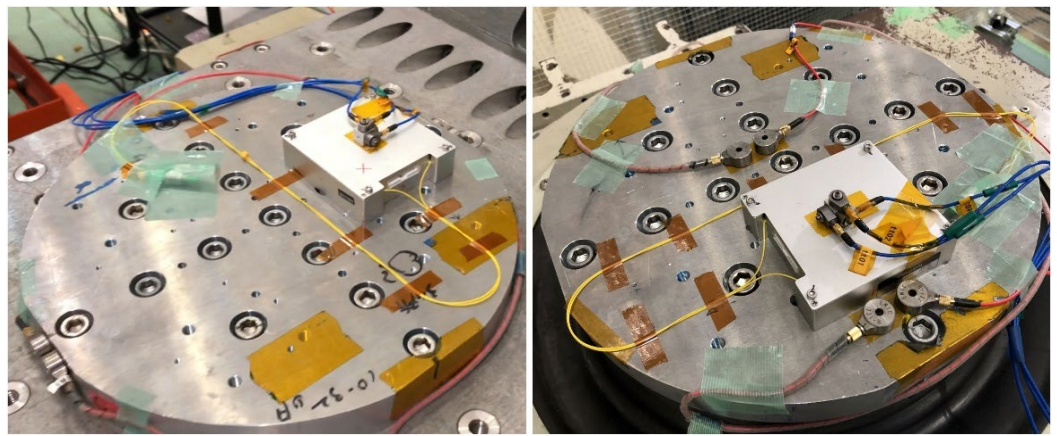

**Figure 7.** EDFA module during vibration tests on the X-Y horizontal vibration bench (left) and Z vertical vibration bench (right).

## 3.3. Thermal vacuum tests

A prototype module of the CubeSat's EDFA was used several months before the final TVT for a preliminary functional long-duration test under high-temperature vacuum (hot cycling). The result of this test can be seen in fig. 8, which shows the dependence of the output optical power on the EDFA's temperature during one hour of continuous operation starting at a temperature of +55°C. The EDFA's typical behavior can be confirmed, with the output power dropping as the temperature rises. Fig. 9 shows a picture of the EDFA's module in the final stage of this test when its case temperature reached about +90°C.

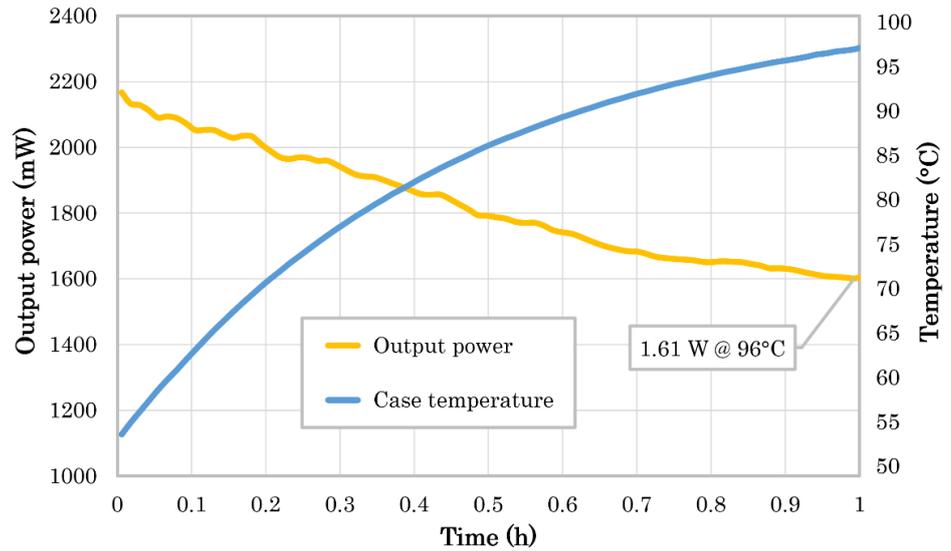

**Figure 8.** EDFA's long-duration high-temperature vacuum test.

The final thermal vacuum tests were performed within a vacuum range of $10^{-6}$ to $10^{-7}$ mbar and a temperature range of -30°C to +55°C with a temperature variation rate of 1°C/min. As shown in fig. 10, multiple functional tests were carried out, with the main emphasis on maintaining a dwell time of 1 hour at extreme levels (-30°C and +55°C) during 3 cycles each. The denomination of the functional tests depends on the cycle temperature and the sequential order. For example, FT1C is the first functional test of the cold cycles and FT3H is the third functional test of the hot cycles. All the functional tests during these extreme levels had a typical duration of 20 minutes.

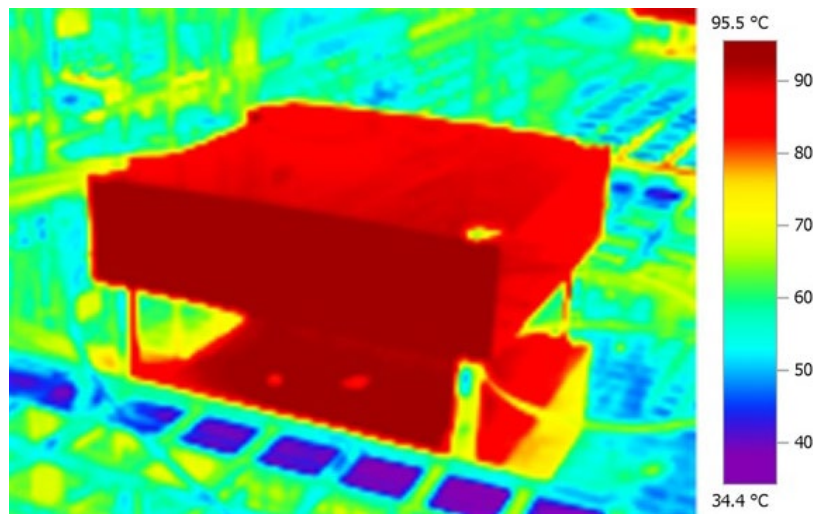

**Figure 9.** EDFA under high-temperature vacuum test.

To monitor the EDFA's case temperature, type-K thermocouples (TC) were attached in the middle point of each case surface except for the top and bottom surfaces which had two TCs in each surface. As shown in fig. 11, the EDFA module was placed inside a jig with heaters in the cube sides. During the hot cycles (+55°C), the case temperature reached a temperature of around +90°C and during the cold cycles (-30°C) around +10°C at the end of the cycles (see fig. 10).

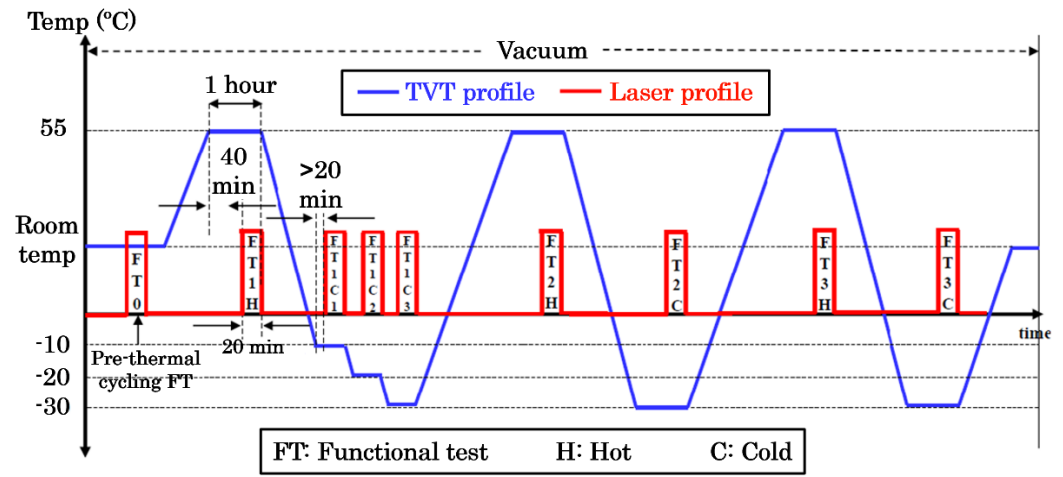

**Figure 10.** Functional tests under thermal vacuum environment.

Figure 12 shows the result of the EDFA's module functional verification during a continuous 24-hour thermal vacuum test including three cold cycles and three hot cycles. The EDFA's operation under extreme-temperature conditions can be understood from the results taken in this measurement campaign. The case temperature rises after the amplifier is enabled, which has a different impact on the optical output power depending on the temperature. The output power tends to fall when the temperature rises during the hot cycles and to rise with temperature during the cold cycles.

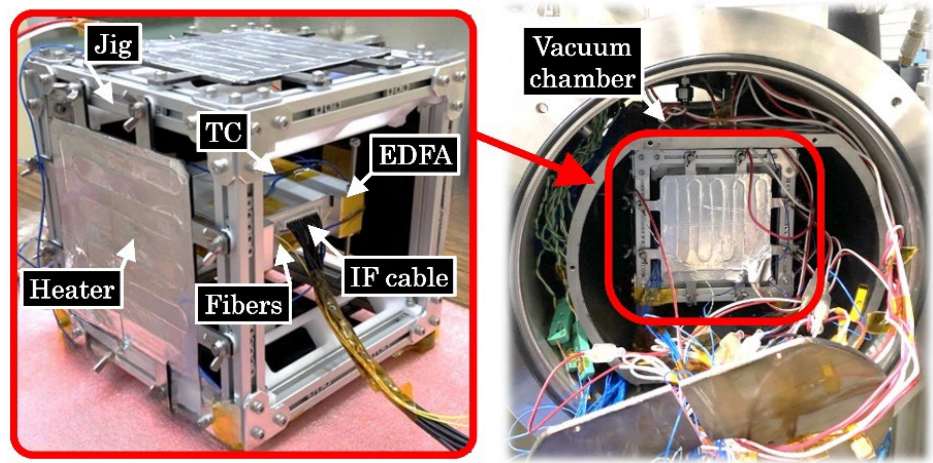

**Figure 11.** EDFA module inside a jig with the heaters and thermocouples for the thermal vacuum tests.

This experiment was designed to achieve the required output optical power of 2 W (33 dBm) during the 20-min functional tests, in the beginning of the hot cycles and in the end of the cold cycles. An initial experiment (FT0) was done under vacuum but in ambient temperature before the thermal cycling to confirm the negligible optical power loss (in the order of 0.1 dB) during the 20-min test. It can be observed that the optical power loss is around 1 dB after the 20-min test during the hot cycles after the EDFA has increased its temperature in about 30ºC to reach close to 90ºC, which was considered the maximum tolerable case temperature to avoid fatal failure. To reach 2 W after the 20-min test during the cold cycles, the initial temperature was set about 3 dB below. Two additional cold cycles (FT1C1 and FT1C2) were done to confirm the smaller power variation with less-extreme cold temperatures. Table 1 shows the detailed numerical result of the full TVT campaign.

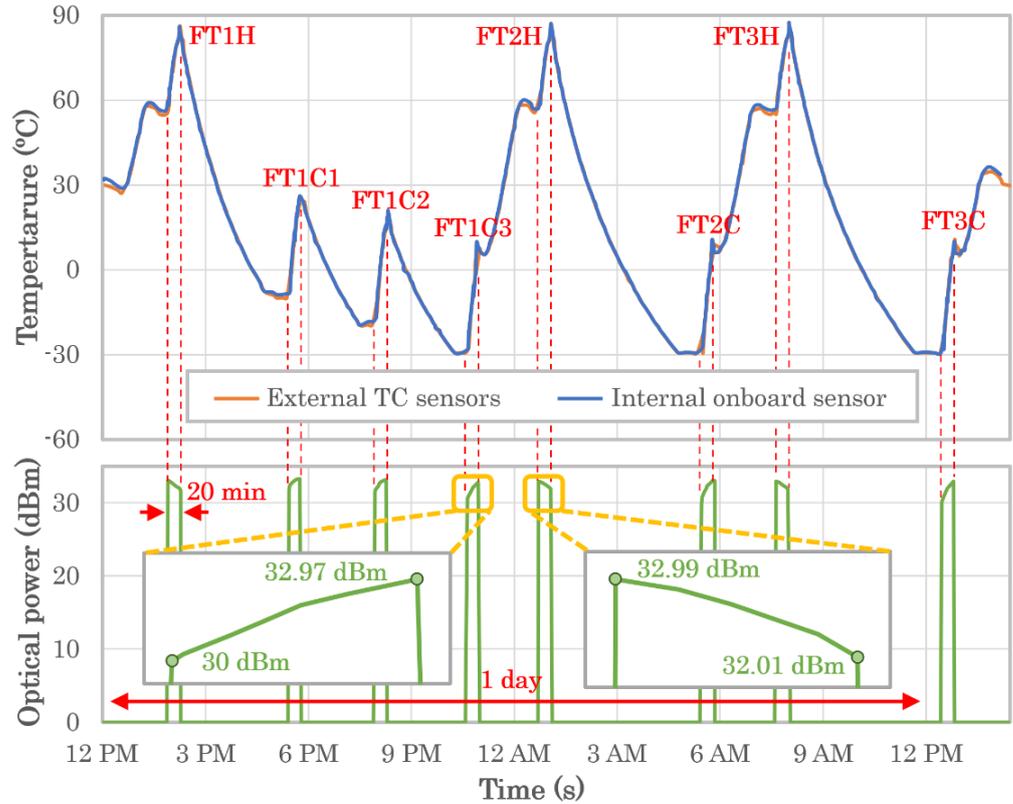

**Figure 12.** EDFA's module full functional verification during 24-hour thermal vacuum test with cold and hot cycles.

The trend that optical power follows with temperature can be identified and characterized depending on the initial temperature before laser pump, hence depending on the operation time, the lasercom-terminal transmitted power can be tuned this way through the EDFA settings in order to satisfy the link-budget requirements. LEO-ground downlinks occur during passes shorter than 10 minutes, and typically shorter than 6 minutes [3], with an average duration of 5 minutes [5]. As the TVT measurement in fig. 7 shows, the effective output power can be guaranteed to be over 2 W during the whole LEO-ground lasercom link by starting to transmit at 2.2 W, which is allowed by the amplifier configuration. Therefore, it can be confirmed that this EDFA is suitable for LEO-ground communications, which is the main application of this amplifier.

However, the 6U-CubeSOTA version (see fig. 1, right) can be applied to intersatellite links from LEO to the geosynchronous orbit (GEO) as well in order to transmit data to the ground through a MEO (medium Earth orbit) or GEO relay in the middle. The main advantage of this application is the much higher link availability because of the longer time under direct line of sight, which on average is 57 minutes between LEO and GEO [5], as opposed to the 5 minutes of LEO-ground. It has been concluded that the CubeSat's EDFA can support this application as well if the starting output power is set to 2.5 W by increasing the pump-laser current, which is the maximum achievable power, reaching a minimum optical power around 2 W after 1 hour. It must be noted that this is the worst case, and in the other cases, where initial temperature is lower and/or operation time is shorter, optical power will be well above 2 W during the full link. Since there is enough time to cool down the EDFA between one LEO-GEO contact and the following one, in practice this amplifier can guarantee continuous operation in this scenario, satisfying the most important merit of LEO-GEO, i.e., its superior link availability.

Table 1. Numerical results of the EDFA's full TVT campaign.

| Functional test name | Initial-final power (dBm) | Consumed power (W) | Initial-final temp. (ºC) |
|---|---|---|---|
| FT0 | 32.82 | 18.58 | Ambient |
| FT1H | 32.03 to 30.99 | 18.67 | 56.5 to 86.9 |
| FT1C1 | 31.6 to 33.2 | 17.92 | -10 to 25 |
| FT1C2 | 31 to 33.12 | 18.52 | -20 to 15 |
| FT1C3 | 30 to 32.97 | 17.99 | -30 to 5 |
| FT2H | 32.99 to 32.01 | 18.71 | 57.6 to 87.6 |
| FT2C | 30 to 32.99 | 17.79 | -30 to 5 |
| FT3H | 31.90 to 30.82 | 18.66 | 56.8 to 88.1 |
| FT3C | 30 to 32.92 | 17.74 | -30 to 5 |

## 4. Conclusions

This manuscript describes a miniaturized EDFA developed and qualified for space operation, with a form factor compatible with the CubeSat platform. The optical power to size ratio is the biggest in the scientific literature and the satellite market to the best of the authors' knowledge [10] [11], meaning that this is the world's smallest space-qualified high-power optical amplifier. NICT plans to test it in orbit as a key component of the CubeSOTA mission, currently in preparation [5]. Although this manuscript has focused on the high-power amplifier of the CubeSOTA terminal, NICT has also conceived its counterpart and it is currently designing and developing a low-noise amplifier to be integrated in the same case to support bidirectional communications.